# Contribution of Graphene Molecules C53 C52 C51 on Astronomical Diffuse Interstellar Bands (DIB)

Norio Ota

Graduate school of Pure and Applied Sciences, University of Tsukuba, *Japan*

*https://www.researchgate.net/profile/Norio-Ota-Or-Ohta*

This molecular orbital analysis predicts that pure carbon graphene molecules would play an important role on astronomically observed Diffuse Interstellar Bands (DIB), rather than fullerene. Laboratory experiments precisely coincided with observed DIB bands as studied by E. Cambell et al., which were considered to originate from mono-cation fullerene-$(C_{60})^{1+}$. To check theoretically a molecular orbital excitation of $(C_{60})^{1+}$ was calculated by the Time-Dependent DFT. Calculated two bands were close to observed DIBs, but there were two problems, that the oscillator strength was zero, and that other three DIBs could not be reproduced. Laboratory experiments was the mass spectroscopic one filtering m/e=724, to suggest fullerene-$(C_{60})^{1+}$ combined with He. However, there were other capabilities, as like He-atom intercalated 3D-graphite, [graphene$(C_{53})^{1+}$ --He--$(C_7)$], [graphene$(C_{51})^{1+}$ --He--$(C_9)$] and so on. A family of graphene $(C_{53})$, $(C_{52})$ and $(C_{51})$ was calculated. Results show that an astronomically observed 957.74nm band was reproduced well by calculated 957.74nm, also confirmed by laboratory experiment of 957.75nm. Other observed 963.26, 936.57 and 934.85nm bands were calculated to be 963.08, 935.89 and 933.72nm. Moreover, experimental 922.27nm band was calculated to be 922.02nm, which is not yet astronomically observed. Similarly, experimental 925.96, 912.80, 909.71 and 908.40nm bands were calculated to 926.01, 912.52, 910.32 and 908.55nm. It should be emphasized that graphene molecules may be ubiquitously floating in interstellar space.

**Key words**: DIB, carbon, graphene, fullerene, TD-DFT

## 1. Introduction

This paper theoretically predicts background molecules of astronomically observed Diffuse Interstellar Bands (DIB)[1]. In our previous papers [2),3)], we pointed out that some observed Infrared spectrum (IR) could be reproduced well by the quantum chemically calculated molecular vibrational infrared spectrum using the Density Functional Theory (DFT). Molecules were polycyclic pure carbon molecules as like soccer ball like fullerene-$(C_{60})$, plane graphene molecules $(C_{53})$, $(C_{52})$, $(C_{51})$, $(C_{23})$, $(C_{22})$ and $(C_{21})$ having one or two pentagon carbon units among hexagon networks. In this study, we like to apply above carbon molecules to reproduce DIB using the Time dependent DFT (TD-DFT) calculation[4), 5)].

In our very recent paper[6)], we had a success that TD-DFT could explain many observed DIBs by some specified Polycyclic Aromatic Hydrocarbon molecules (PAHs) as like $(C_{53}H_{18})$, and $(C_{23}H_{12})$. Those molecules were selected under the astronomical top-down material creation model[7)], could reproduce well-known astronomical PAH bands in a wide wavelength from 3.2, 6.3, 7.7, 8.6, 11.2, to 12.7 micrometer[8)].

The Diffuse Interstellar Bands (DIB) are a large set of absorption features at a wide range of visible to infrared wavelengths to associate with pure carbon and /or hydrocarbon molecules as reviewed by T. R. Geballe[1)]. The discovery of the first diffuse interstellar bands was made by Mary Lea Heger[9)], almost 100 years ago. Until now, there observed over 500 DIBs[10)-15)]. However, any specific molecule had not been identified yet. After long years efforts[16)-19)], five near-infrared DIBs were matched to the laboratory experimental spectrum by singly cationic fullerene $(C_{60})^{1+}$ [20),21)] in 2015-2016 by E. Cambell et al. However, other many DIBs were not identified yet. This is the great mystery for the astrochemistry and molecular science. It was believed that DIB may come from molecular orbital photoexcitation as the absorbed bands. In this study, to obtain molecular excitation energy, we should also apply TD -DFT calculation.

Purpose of this study is to indicate the specific pure carbon molecules to satisfy both astronomically observed IR and DIB.

## 2. Calculation Methods

In calculation, we used DFT[21,22] and TD-DFT[5] with the unrestricted PBEPBE functional[23), 24]. We utilized the Gaussian09 software package[25] employing an atomic orbital 6-31G basis set[26]. Unrestricted DFT calculation was done to have the spin dependent atomic structure. The required convergence of the root-mean-square density matrix was $10^{-8}$. Based on such optimized molecular configuration, fundamental vibrational modes were calculated. This calculation also gives harmonic vibrational frequency and intensity in infrared region. To each IR spectral line, we assigned a Gaussian profile with a full width at half maximum (FWHM) of $4cm^{-1}$.

For obtaining the molecular orbital excitation energy and the oscillating strength, we tried Time-Dependent DFT calculation in Gaussian09 package. Excited energies were calculated simultaneously up to 14th excitations.

The scaling factor "s" between the experimental energy to the DFT and TD-DFT calculated energy is s=1.010, estimated from fundamental molecular vibrational modes of fullerene-($C_{60}$) compared with laboratory experiments to match with largest intensity IR band at 18.82 micrometer.

## 3. Carbon Pentagon-Hexagon Combined Graphene Molecules

In our previous papers [2),3], astronomically observed carbon related IR were reproduced well by the carbon pentagon-hexagon combined molecules. Those may be ubiquitously floating in interstellar space. We can suppose major contribution on DIB. As shown in Fig. 1, their molecular features were illustrated. Fully hexagon networked molecule ($C_{54}$) shows stable spin state of *Sz*=0/2. Floating ($C_{54}$) would be attacked by high-speed proton or electron coming from the central star as like our solar system, kicked out one carbon atom transforming from ($C_{54}$) to ($C_{53}$) having one pentagon site. Circled number ⑤ shows the pentagon site. Also, high energy photon irradiated from the central star may pull out one electron to change neutral molecule to cation. Stable spin-state of ($C_{53}$) was *Sz*=2/2. It's spin-cloud was illustrated in right column, where up-spin was drawn by red, down-spin by blue. Successively, attack by proton/electron creates ($C_{52}$), ($C_{51}$) and so on. Side view of those molecules varies from plane to cup like configuration. Finally, we can expect the creation of soccer ball like fullerene-($C_{60}$). Such a molecular creation system was named by the astronomical top-down scheme.

Stable spin state of cationic molecule was calculated. For fullerene-$(C_{60})^{1+}$, graphene-$(C_{53})^{1+}$, $(C_{52})^{1+}$ and $(C_{51})^{1+}$, spin state was *Sz*=1/2, while for $(C_{53})^{2+}$ high spin state of *Sz*=4/2.

## 4. Molecular Orbitals and Excitation Energy

Typical examples of calculated molecular orbital diagram by TD-DFT calculation were illustrated in Fig. 2, for mono cation fullerene $(C_{60})^{1+}$ and for mono cation graphene $(C_{53})^{1+}$. Optimization of molecular configuration and spin density was previously done by DFT. Energy level of Alpha orbit was marked by blue bar, while Beta orbit by red.

In case of mono-cation fullerene-$(C_{60})^{1+}$, Alpha-HOMO orbit was number 180 orbit (#180), while Beta one was #179. We obtained excitation energy among those molecular orbits until 14th excitation. Lowest one was absorption of N1 from Beta#176 to #180 by wavelength of N1=15174nm, very long wavelength. According to excitation number, wavelength would become shorter (higher energy). Detailed results were listed in Table 2. It is interesting that N9 (Beta #171 to Beta LUMO #180) shows 954nm, which is close to observed DIB. Also, N11 (from #167 to #180) was 936nm, again close to DIB. Detailed comparison will be discussed in later.

Mono-cation graphene $(C_{53})^{1+}$ show the excitation energy N1 (from Beta-HOMO #158 to Beta-LUMO #159) to be 3898nm. As listed in Table 3, N8 (from Alpha-HOMO #159 to #161) was 957nm, N10 (from #157 to Alpha-HOMO #160) was 936nm. Both could well reproduce observed DIB. Detailed comparison will be discussed in later.

In Fig. 3, we compared reported observed bands with calculated bands of fullerene-$(C_{60})^{1+}$ and graphene-$(C_{53})^{1+}$. Reports were referenced in this paper[16)-20]. Upper left blue column was edited DIBs by T. Lan et al.[37] in 2015 based on the list by P. Jenniskens et al. [38] in 1994, which bands were related to hydrocarbon molecules as studied in our recent paper[6]. In case of pure carbon, longer wavelength region from 900nm to some micrometer will be main playing field. Especially, estimated fullerene-$(C_{60})^{1+}$ region (920 to 970nm) should be studied in detail.

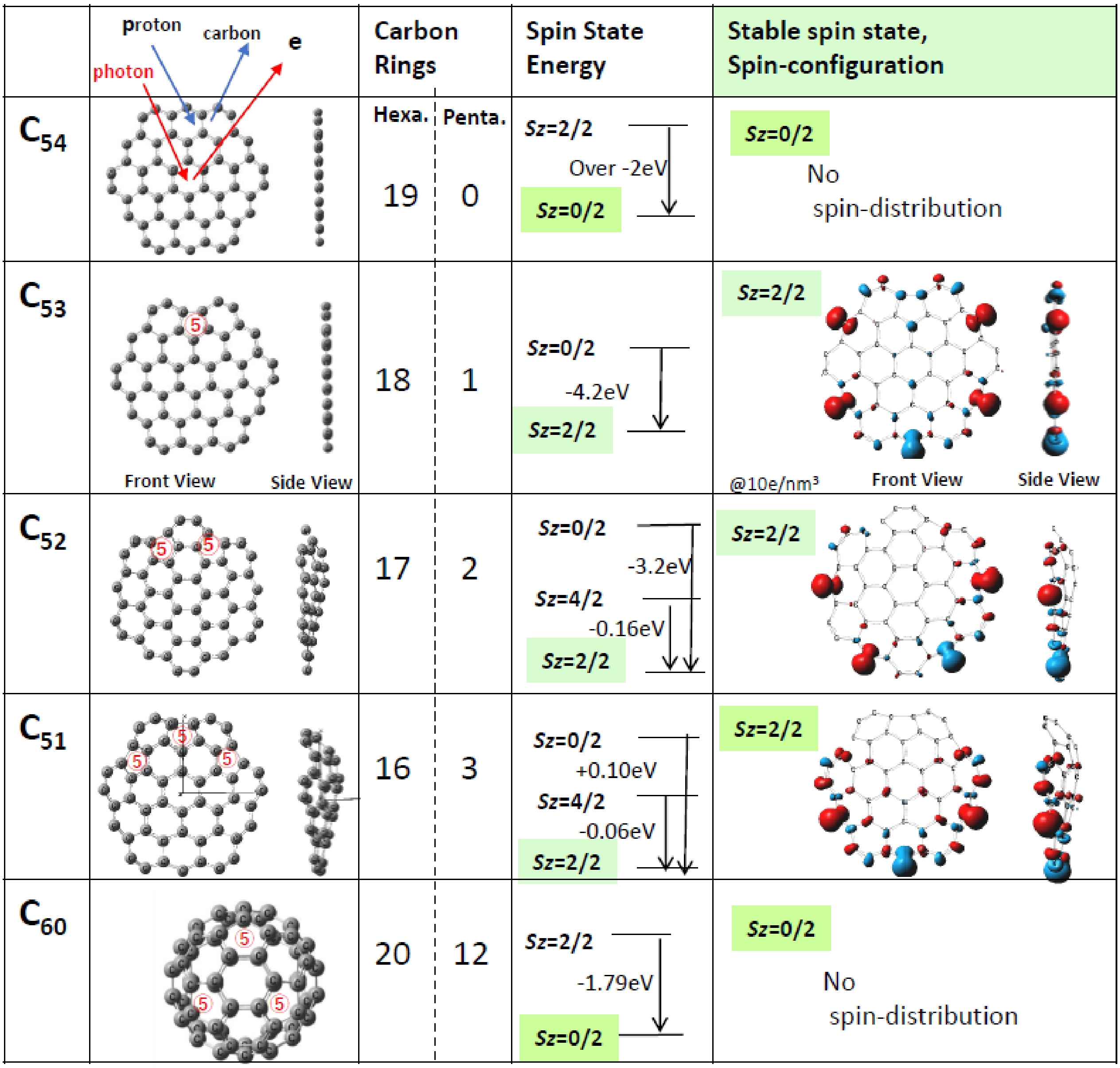

**Fig. 1** Carbon pentagon-hexagon combined molecules. Molecular stable configuration, Energy dependence on spin-state, Spin cloud map for up-spin by red, down-spin by blue. Circled ⑤ shows pentagon site.

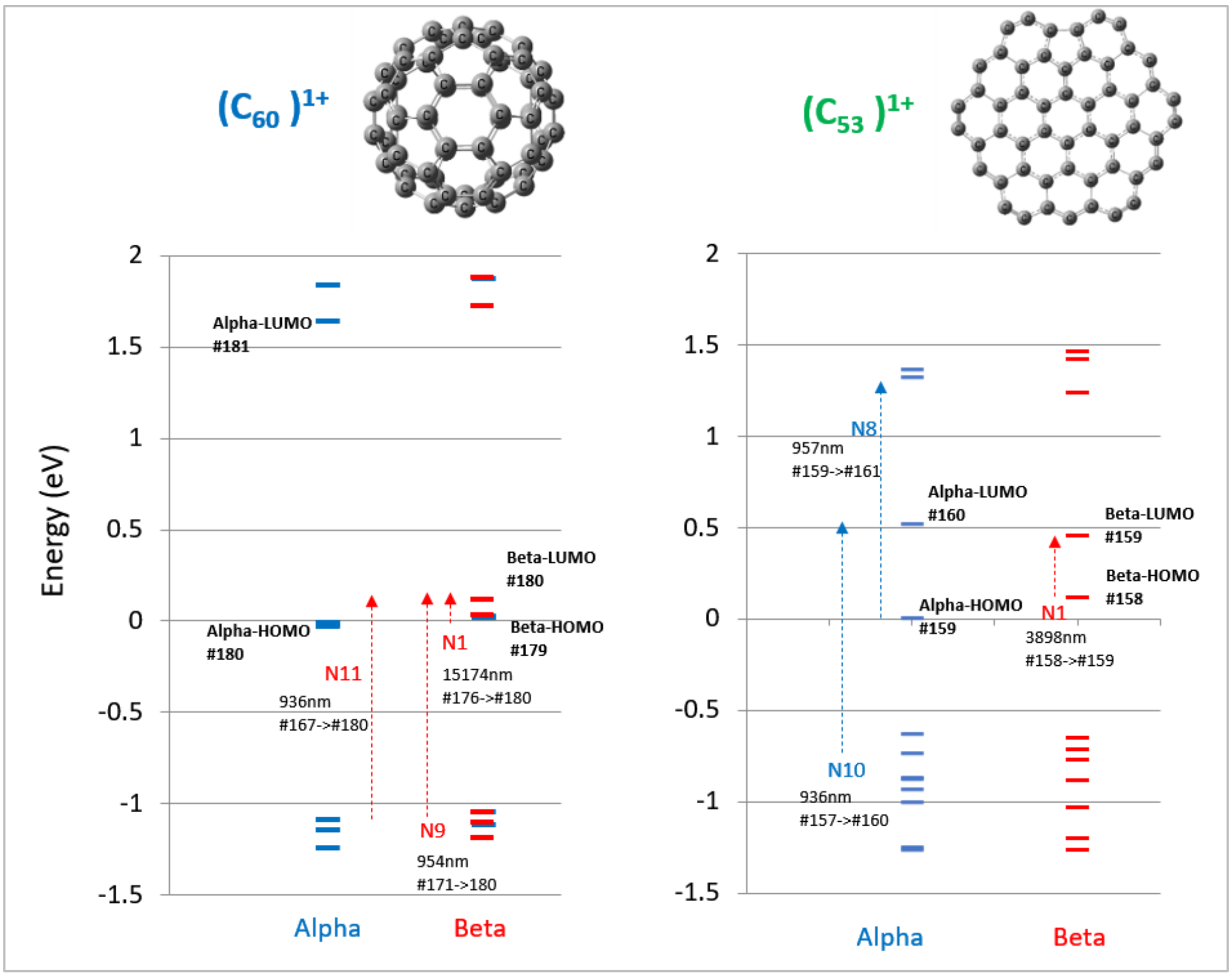

**Fig. 2.** Molecular orbital and excitation energy for mono cation fullerene $(C_{60})^{1+}$ and graphene $(C_{53})^{1+}$.

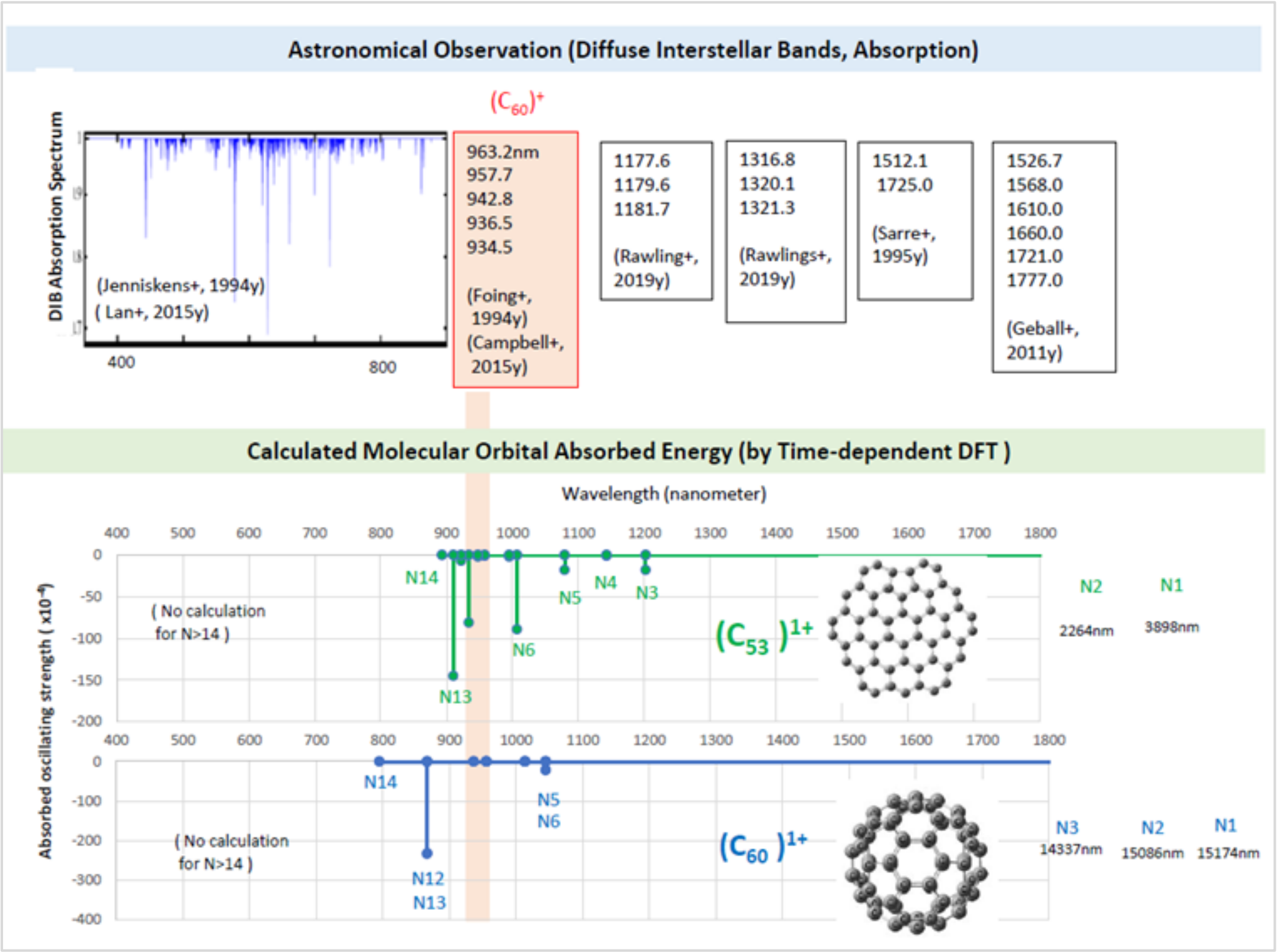

**Fig. 3** Comparison of observed DIBs with calculated excitation wavelength (nm) of $(C_{60})^{1+}$ and $(C_{53})^{1+}$.

## 5. Theoretical Approach to Fullerene-$(C_{60})^{1+}$

Precise laboratory experiment to identify observed DIB was done by E. Campbell et al. in 2015 for the first time predicting specific molecule[19]. As noted in Table 1, DIB9632, named by observed band central wavelength of 9632.0Angstrom, 963.20nm, coincided with experimental 963.27nm, also DIB9577 (observed 957.74nm) to 957.75nm, DIB9428 (942,84nm) to 942.85nm, DIB9365 (936.57nm) to 936.59nm, and additionally DIB9345 (934.85nm) to 934.91nm in 2016[20]. Coincidence was excellent. Fig. 5 shows complete coincidence between observed DIB (blue dotted vertical line) and experiment (red points). Moreover, they add extra experiments of shorter wavelength absorbed bands, which are not yet astronomically observed bands. They lie on red broken line in the figure, just y/x=1 line.

Background molecule was believed to be mono-cation fullerene-$(C_{60})^{1+}$. Experiment was done in He gas phase at low temperature 6K to simulate well interstellar environment. Mass-spectroscopic filtering was done to set a window of (m/e=724). Mono-cation fullerene $(C_{60})^{1+}$ coupled with atomic-He, noted as [fullerene-$(C_{60})^{1+}$--He] passes this window.

In this study, we like to theoretically confirm above experiments. Table-2 is TD-DFT calculated results. Neutral fullerene-$(C_{60})$ show calculated N5=652.69nm coincide with DIB at 653.21nm within observed FWHM=1.72nm[38]. Similarly, eleven calculated bands from N6 to N14 show reasonable coincidence. However, there is a problem that calculated oscillator strength of those bands were all zero, which may come from high symmetrical configuration of soccer ball like fullerene. Excited electron cloud would be polarized, but be cancelled out by high symmetrical oppositely polarized excitation.

We also checked mono-cation fullerene-$(C_{60})^{1+}$ as noted in Table-2. Calculated N11=936.10nm was almost coincident with observed DIB9365 but remain uncertainty. Theoretical carefulness is necessary. We tuned the scaling factor by three cases of s=1.013, 1.010, and 1.007. Results were compared in Table 1 and Fig. 6. In case of s=1.010, N11=936.10nm is close to observed DIB9365 (observed wavelength 936.57nm). While in case of s=1.007, calculated 938.89nm was far from observation. N9 shows calculated 955.23nm for s=1.010, which was far from DIB9577 (957.74nm), but for s=1.007, calculated 958.08nm is closer to DIB9577.

By applying the model of fullerene-$(C_{60})^{1+}$, there are two theoretical problems, that calculated oscillator strength is "zero", and that we could not reproduce other four observed DIBs.

## 6. Molecule Candidates for Mass Spectroscopic (m/e=724) Window

Theoretically, fullerene-$(C_{60})^{1+}$ could not show fair coincidence both with astronomical observation and experiments. We should consider actual situation of experiment. Mass spectroscopic molecular filtering means two issues, one is total mass to be 724 (carbon mass=60x12, He mass=4), another is total charge to be (+1). As illustrated on panel (a) of Fig. 4, a simple candidate is [fullerene-$(C_{60})^{1+}$--He], where He are combined with fullerene by Van der Waals attractive force.

We should try other models. Another candidate would be [graphene-$(C_{60})^{1+}$--He] as imaged on (b). Experimental raw carbon material may include bulk 3D-graphite (not 2D-graphene), which will be broken to tiny 3D-graphite by heating. It is interesting that broken 3D-graphite may make He intercalated material as imaged on (c). Typical example is [graphene-$(C_{53})^{1+}$--He--$(C_7)$]. Sometimes, charging process may allow us an example of [graphene-$(C_{53})^{2+}$--He--$(C_7)^{1-}$]. Not only experiments, but also even in interstellar space, there would be many candidates as like [graphene-$(C_{51})^{1+}$--He—$(C_9)$] and so on.

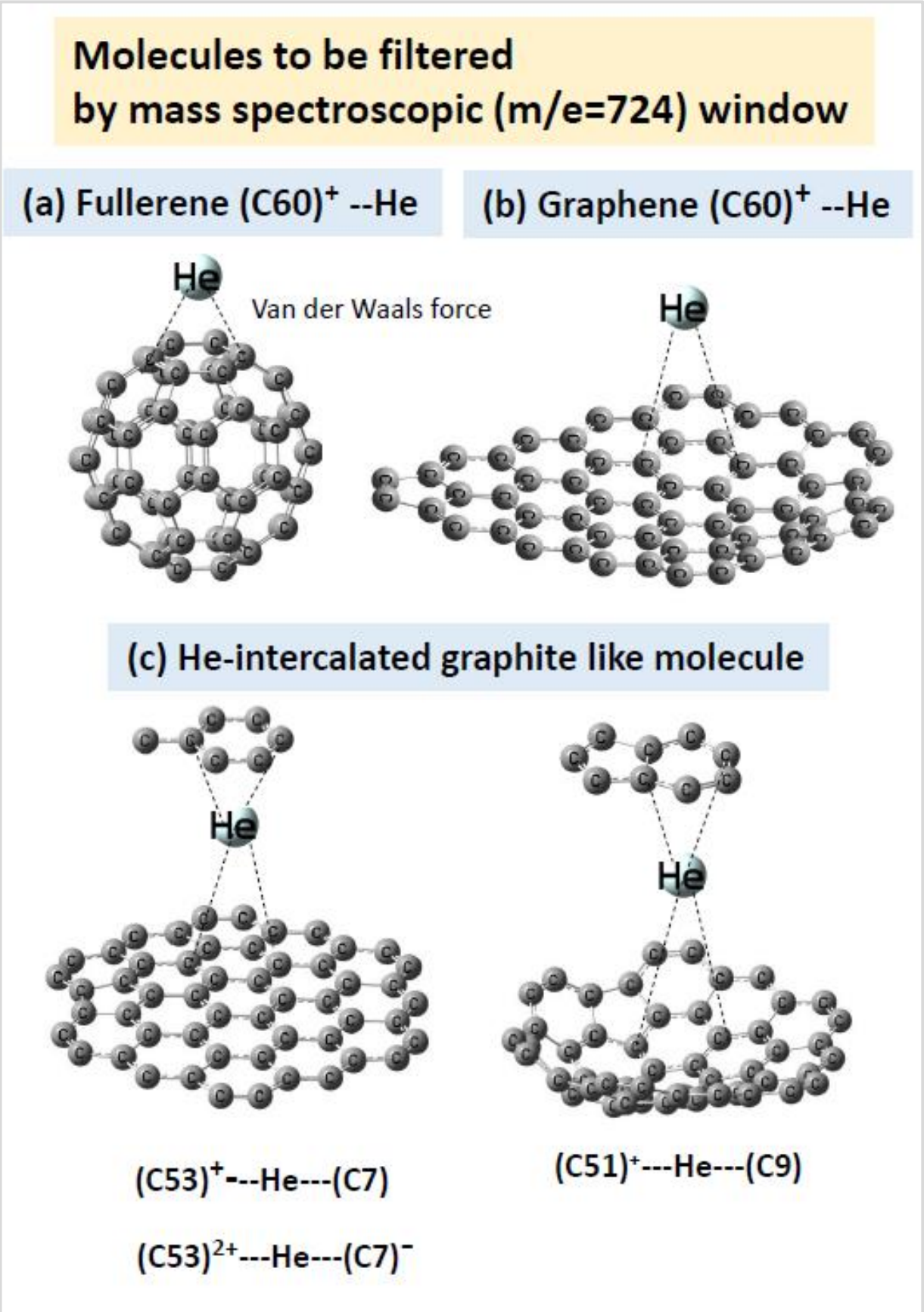

**Fig. 4** Molecular candidates for mass spectroscopic filtering of (m/e=724).

**Table 1.** Pure carbon related astronomical DIB, laboratory experiments and TD-DFT calculated molecular orbital excitation wavelength.

Unit: nm (wavelength in nanometer)

| DIB name | | | | | | | | DIB9345 | DIB9365 | DIB9428 | DIB9577 | DIB9632 |
|---|---|---|---|---|---|---|---|---|---|---|---|---|
| Astronomical Observation | A. Webster (1993), Jenniskens (1997) | | | | | | | | | | | 963.20 |
| | G. Walker (2015) | | | | | | | | 936.57 | 942.84 | 957.74 | |
| | E. Campbell (2016) | | | | | | | 934.85 | | | | |
| Laboratory Experiment | (C60)+ by E. Campbell (2015) (2016) | | 908.43 | 909.71 | 912.80 | 922.27 | 925.96 | 934.91 | 936.59 | 942.85 | 957.75 | 963.27 |
| | | Scaling factor | | | | | | | | | | |
| Quantum Calculation (This Study) | | s=1.013 | | | | | | | 933.33 | | 952.40 | |
| | Fullerene (C60)+ | s=1.010 | | | | | | | **936.10** | | 955.23 | |
| | | s=1.007 | | | | | | | 938.89 | | **958.08** | |
| | | s=1.013 | | 907.63 | | 919.29 | | 930.96 | 933.11 | 944.81 | 954.91 | |
| | (C53)+, Sz=1/2 | s=1.010 | | **910.32** | | **922.02** | | **933.72** | **935.89** | 947.62 | **957.74** | |
| | | s=1.007 | | 913.03 | | 924.77 | | 936.50 | 938.68 | 950.44 | 960.594 | |
| | | s=1.013 | 905.86 | | 909.81 | | | | | | 952.23 | 960.23 |
| | (C53)2+, Sz=4/2 | s=1.010 | **908.55** | | **912.52** | | | | | | 955.06 | **963.08** |
| | | s=1.007 | 911.26 | | 915.23 | | | | | | **957.91** | 965.94 |
| | | s=1.013 | | | | | 923.26 | | | | 956.58 | |
| | (C51)+, Sz=1/2 | s=1.010 | | | | | **926.01** | | | | **959.42** | |
| | | s=1.007 | | | | | 928.77 | | | | 962.28 | |

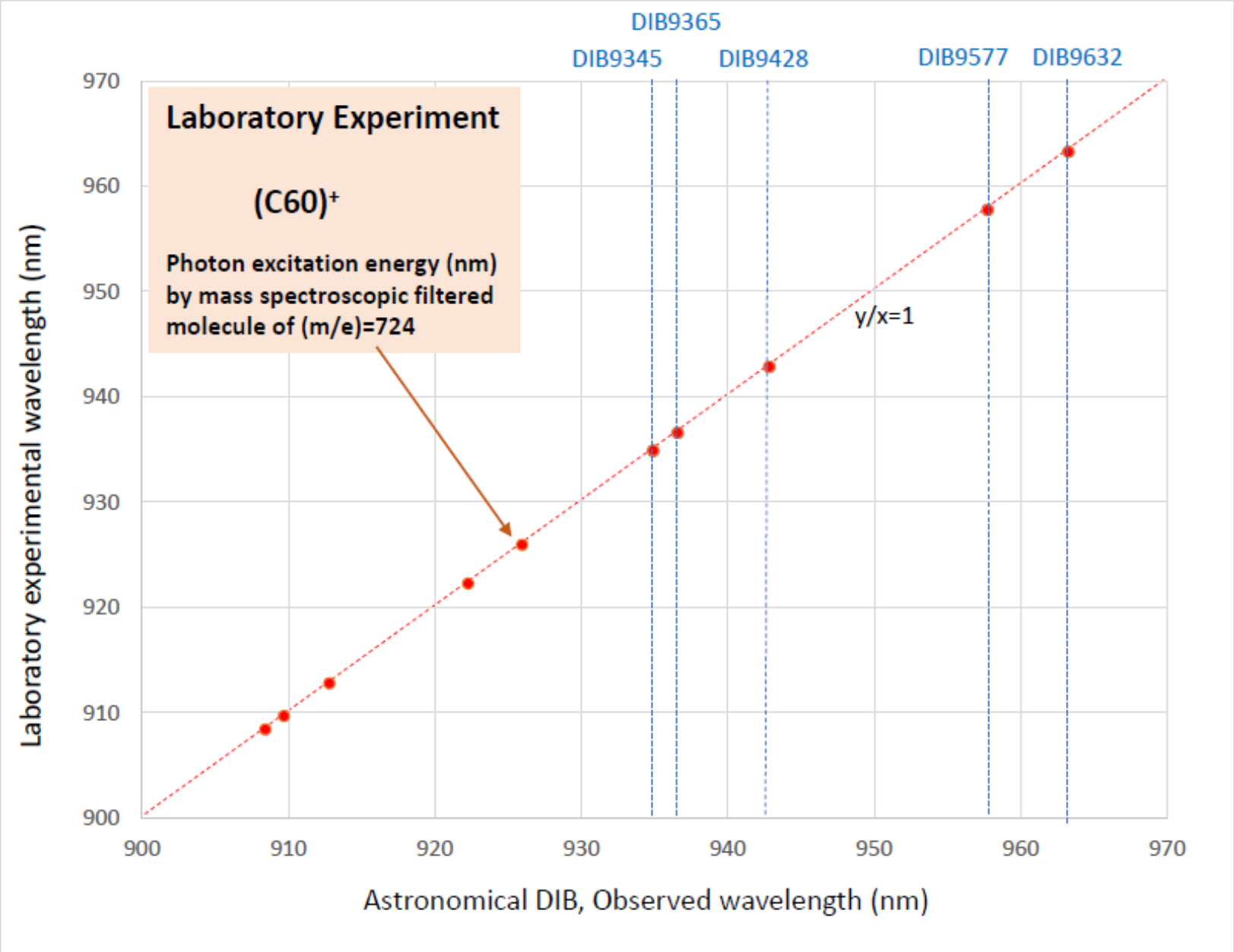

**Fig. 5** Mass spectroscopic experiment by Campbell et al.[19),20)] (red point) precisely coincide with observed DIBs (blue dotted vertical lines).

**Table 2.** TD-DFT calculated molecular excitation wavelength for a charge neutral fullerene-($C_{60}$) and mono-cation fullerene-$(C_{60})^{1+}$ compared with astronomical DIB and laboratory experiments. Yellow marked column is coincided wavelength. In a DIB column, "x" means out of observed DIB list.

| TD-DFT Calculation | | | | | DIB, Astronomical Observation | | | Laboratory experiment | |
|---|---|---|---|---|---|---|---|---|---|
| Molecule | Excitation number | Calculated Wavelength (nm) s=1.010 | Oscillator strength (10E−4) | | Observed Wavelength (nm) | FWHM (nm) | | Experimental Wavelength(nm) | FWHM (nm) |
| C60 c0 m0 | N1 | 669.87 | 0 | | 669.93 | 0.12 | | | |
| | N2 | 669.76 | 0 | | 669.93 | 0.12 | | | |
| | N3 | 669.71 | 0 | | 669.93 | 0.12 | | | |
| | N4 | 669.67 | 0 | | 669.93 | 0.12 | | | |
| | N5 | 652.69 | 0 | | 653.21 | 1.72 | | | |
| | N6 | 652.59 | 0 | | 653.21 | 1.72 | | | |
| | N7 | 652.48 | 0 | | 653.21 | 1.72 | | | |
| | N8 | 644.93 | 0 | | 645.16 | 2.54 | | | |
| $(C60)^0$, | N9 | 644.90 | 0 | | 645.16 | 2.54 | | | |
| Sz=0/2 | N10 | 644.58 | 0 | | 645.16 | 2.54 | | | |
| | N11 | 619.01 | 0 | | 617.72 | 2.30 | | | |
| | N12 | 618.99 | 0 | | 617.72 | 2.30 | | | |
| | N13 | 618.84 | 0 | | 617.72 | 2.30 | | | |
| | N14 | 618.73 | 0 | | 617.72 | 2.30 | | | |
| | N14 | 794.43 | 0 | | 792.78 | 1.50 | | | |
| C60 c1 m1 | N1 | 15174.85 | 0 | | x | | | | |
| | N2 | 15086.83 | 0 | | x | | | | |
| | N3 | 14337.92 | 0 | | x | | | | |
| | N4 | 14231.98 | 0 | | x | | | | |
| | N5 | 1044.22 | −21 | | x | | | | |
| | N6 | 1044.06 | −22 | | x | | | | |
| | N7 | 1013.30 | 0 | | x | | | | |
| $(C60)^{1+}$, | N8 | 1012.88 | 0 | | x | | | | |
| Sz=1/2 | N9 | 955.23 | 0 | | 957.74 | 0.33 | strong | 957.75 | 0.25 |
| | N10 | 954.96 | 0 | | x | | | | |
| | N11 | 936.10 | 0 | | 936.57 | 0.25 | weak | 936.59 | 0.24 |
| | N12 | 866.68 | −232 | | 864.82 | 0.41 | | | |
| | N13 | 865.99 | −234 | | 864.82 | 0.41 | | | |
| | N14 | 794.43 | 0 | | 792.78 | 1.50 | | | |

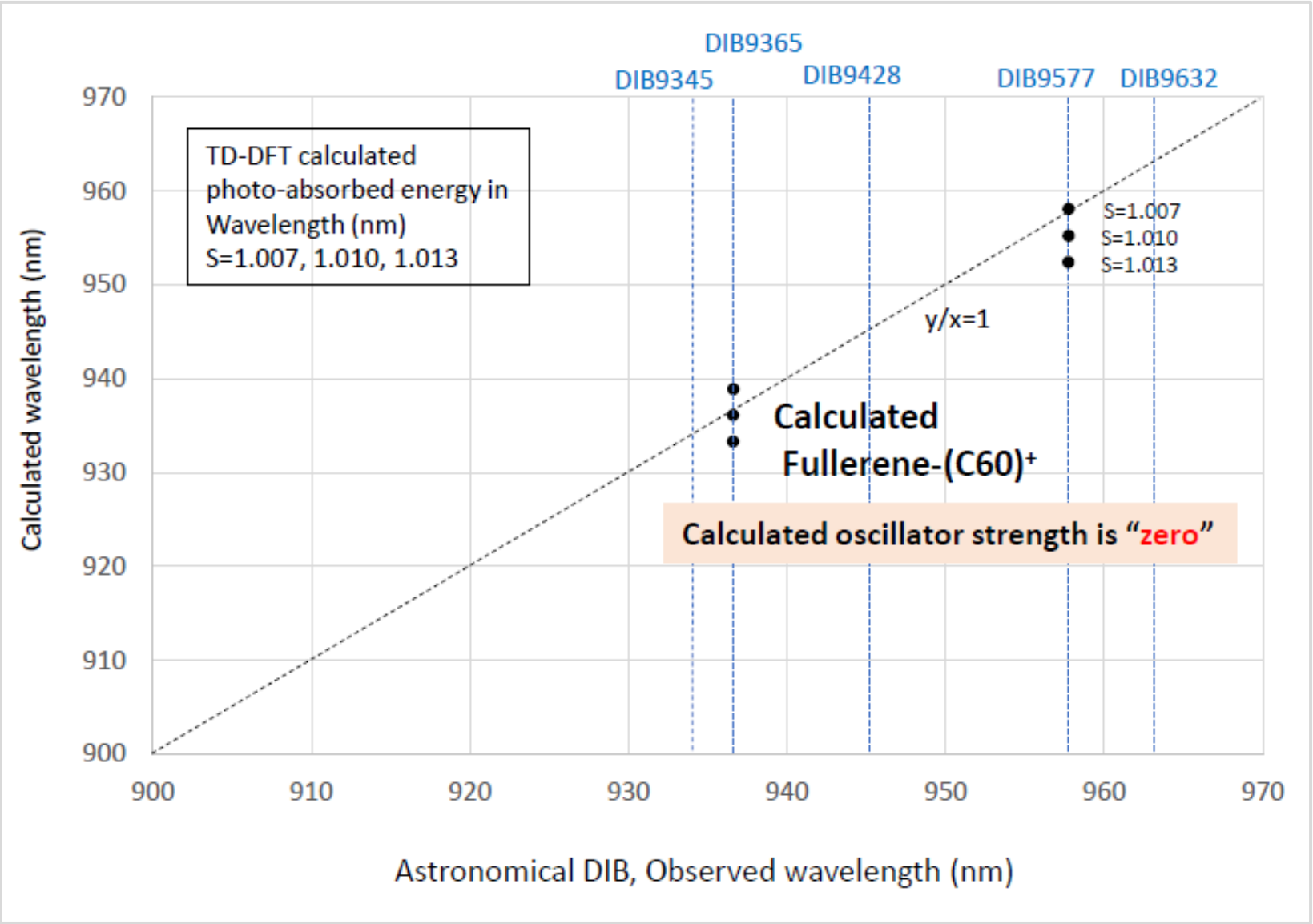

**Fig. 6** Calculated molecular orbital excitation wavelength for fullerene-$(C_{60})^{1+}$ compared with astronomical DIBs (blue dotted lines). Calculated value varies with the scaling factor "s", for s=1.007, 1.010, 1.013.

## 7. Graphene Molecules

Molecular orbital excitation wavelength was calculated on a series of graphene molecules as like ($C_{53}$), ($C_{53}$)$^{1+}$, ($C_{53}$)$^{2+}$, ($C_{52}$) and ($C_{51}$), which detailed results are listed in Table 3, where coincide band both with astronomical DIB and laboratory experiments are marked by yellow column. Coincided band with laboratory experiments are marked by light red column.

It was a surprise that the excitation number N8 of graphene-($C_{53}$)$^{1+}$ show complete coincidence both with astronomical DIB of 957.74nm, and with experimental one of 957.75nm. Also, we recognize close coincidence by N10=935.89nm and N11=933.72nm with DIB bands of 936.57nm and 934.85nm respectively.

Concerning di-cation molecule ($C_{53}$)$^{2+}$, we could recognize good coincidence by N10=963.08nm compared with DIB of 963.20nm, which are experimentally confirmed to be 963.27nm within FWHM=0.22nm. Also, N11=955.06nm presents close value to DIB 957.74nm. Moreover, three carbon pentagons included molecule graphene-($C_{51}$)$^{1+}$ shows N3=959.42nm fairly close to DIB of 957.74nm and to experimental 957.75nm.

Moreover, it should be noted that experimental 922.27nm band was calculated to be 922.02nm. Similarly, experimental 925.96, 912.80, 909.71 and 908.40nm bands were all reproduced well by calculated 926.01, 912.52, 910.32 and 908.55nm as compared in Table 1.

We need very carefull theoretical comparison on the scaling factor "s". In Fig. 7, we illustrated variation on the scaling factor. Scaling factor s=1.007 gives longer wavelength than s=1.010, while s=1013 shorter wavelength. Difference of molecular species are marked by colored character. We can see good results mostly by the case of s=1.010.

In Fig. 8, we could simply compare calculated values by s=1.010 with astronomical DIBs by vertical blue dotted lines, with laboratory experimented ones by red circles, of which vertical axis is shifted for easy comparison. It is amazing that those three different approaches show good agreement by together.

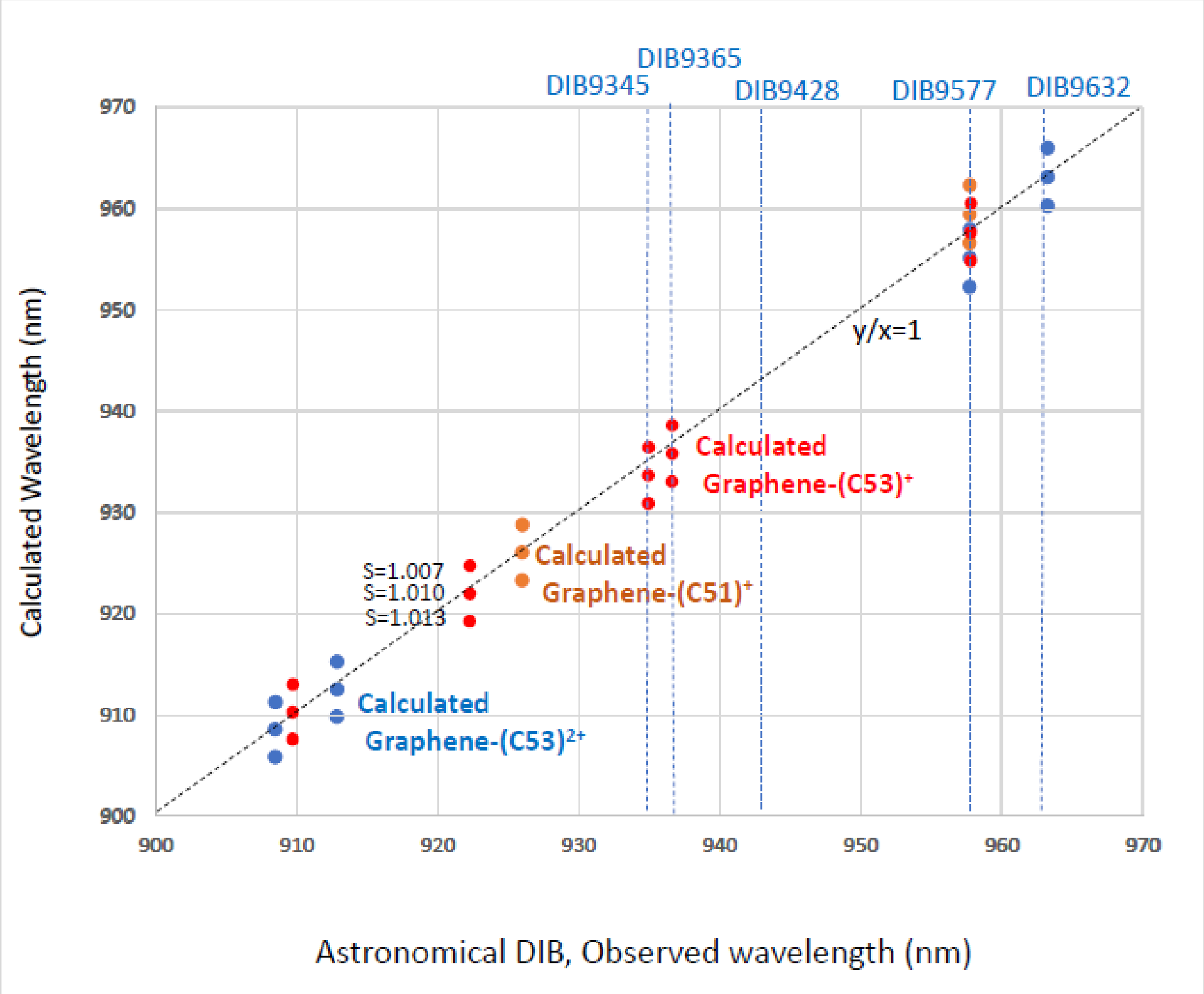

**Fig. 7** Comparison of calculated results of graphene molecules with astronomical DIBs, varying the scaling factor "s".

**Table 3.** Calculated graphene family compared with astronomical DIB and laboratory experiments. Theoretical scaling factor “s” is 1.010.

| TD-DFT Calculation | | | | | DIB, Astronomical Observation | | | Laboratory experiment | |
|---|---|---|---|---|---|---|---|---|---|
| Excitation Number | | Calculated Wavelength (nm) S=1.010 | Oscillator strength (10E-4) | | Observed Wavelength (nm) | FWHM (nm) | | Experimental Wavelength (nm) | FWHM (nm) |
| C53 c0 m2 | N1 | 2820.17 | -17 | | x | | | | |
| | N2 | 2037.08 | 0 | | x | | | | |
| | N3 | 1692.38 | -2 | | x | | | | |
| | N4 | 1449.89 | -1 | | x | | | | |
| | N5 | 1388.82 | -1 | | x | | | | |
| $(C53)^{0}$, | N6 | 1356.18 | 0 | | x | | | | |
| | N7 | 1214.20 | -1 | | x | | | | |
| Sz=2/2 | N8 | 1137.99 | -1 | | x | | | | |
| | N9 | 1102.45 | -2 | | x | | | | |
| | N10 | 1053.64 | -1 | | x | | | | |
| | N11 | 996.59 | -105 | | x | | | | |
| | N12 | 982.91 | -30 | | x | | | | |
| | N13 | 894.34 | 0 | | x | | | | |
| | N14 | 855.01 | -135 | | n | | | | |
| C53 c1 m1 | N1 | 3898.17 | -9 | | x | | | | |
| | N2 | 2264.44 | -47 | | x | | | | |
| | N3 | 1201.80 | -18 | | x | | | | |
| | N4 | 1142.65 | -1 | | x | | | | |
| | N5 | 1079.12 | -18 | | x | | | | |
| $(C53)^{1+}$, | N6 | 1006.81 | -89 | | x | | | | |
| Sz=1/2 | N7 | 994.84 | -2 | | x | | | | |
| | N8 | 957.74 | -1 | | 957.74 | 0.33 | strong | 957.75 | 0.25 |
| | N9 | 947.62 | -2 | | x | | | | |
| | N10 | 935.89 | -81 | | 936.57 | 0.25 | weak | 936.59 | 0.24 |
| | N11 | 933.72 | -29 | | 934.85 | 0.14 | weak | 934.91 | 0.23 |
| | N12 | 922.02 | -7 | | x | | | 922.27 | 0.27 |
| | N13 | 910.32 | -145 | | x | | | 909.71 | 0.22 |
| | N14 | 893.13 | 0 | | x | | | | |
| C53 c2 m4 | N1 | 2798.20 | -1 | | x | | | | |
| | N2 | 1368.00 | -1 | | x | | | | |
| | N3 | 1333.14 | -56 | | x | | | | |
| | N4 | 1251.60 | -9 | | x | | | | |
| | N5 | 1145.28 | -7 | | x | | | | |
| $(C53)^{2+}$, | N6 | 1085.37 | -38 | | x | | | | |
| Sz=4/2 | N7 | 1074.40 | -361 | | x | | | | |
| | N8 | 1023.34 | 0 | | x | | | | |
| | N9 | 1005.94 | -2 | | x | | | | |
| | N10 | 963.08 | -479 | | 963.20 | 0.30 | strong | 963.27 | 0.22 |
| | N11 | 955.06 | -78 | | 957.74 | 0.33 | strong | 957.75 | 0.25 |
| | N12 | 950.55 | -81 | | x | | | | |
| | N13 | 912.52 | 0 | | x | | | 912.80 | 0.30 |
| | N14 | 908.55 | -3 | | x | | | 908.43 | 0.25 |
| C52 c1 m1 | N1 | 2545.51 | -9 | | x | | | | |
| | N2 | 1573.41 | -2 | | x | | | | |
| | N3 | 1251.04 | -2 | | x | | | | |
| | N4 | 1153.35 | -70 | | x | | | | |
| | N5 | 1081.14 | -25 | | x | | | | |
| $(C52)^{1+}$, | N6 | 1028.21 | -309 | | x | | | | |
| Sz=1/2 | N7 | 886.57 | -12 | | x | | | | |
| | N8 | 879.24 | -91 | | x | | | | |
| | N9 | 862.64 | -3 | | x | | | | |
| | N10 | 844.24 | -77 | | x | | | | |
| | N11 | 826.40 | -1 | | 828.30 | 0.12 | | | |
| | N12 | 803.41 | -60 | | 803.70 | 0.17 | | | |
| | N13 | 774.72 | -1 | | 774.82 | 0.45 | | | |
| | N14 | 754.15 | -58 | | 755.85 | 0.15 | | | |
| C51 c1 m1 | N1 | 1319.26 | -8 | | x | | | | |
| | N2 | 1046.25 | -3 | | x | | | | |
| | N3 | 959.42 | -6 | | 957.74 | 0.33 | | 957.75 | 0.25 |
| | N4 | 926.01 | -4 | | x | | | 925.96 | 0.27 |
| | N5 | 886.14 | -3 | | x | | | | |
| $(C51)^{1+}$, | N6 | 866.21 | -11 | | 864.82 | 0.41 | | | |
| Sz=1/2 | N7 | 850.44 | -113 | | n | | | | |
| | N8 | 837.26 | -1 | | n | | | | |
| | N9 | 830.78 | -1 | | n | | | | |
| | N10 | 812.29 | -386 | | n | | | | |
| | N11 | 800.08 | -89 | | n | | | | |
| | N12 | 769.59 | -29 | | 770.97 | 3.35 | | | |
| | N13 | 752.24 | -4 | | n | | | | |
| | N14 | 744.34 | 0 | | 743.21 | 2.22 | | | |

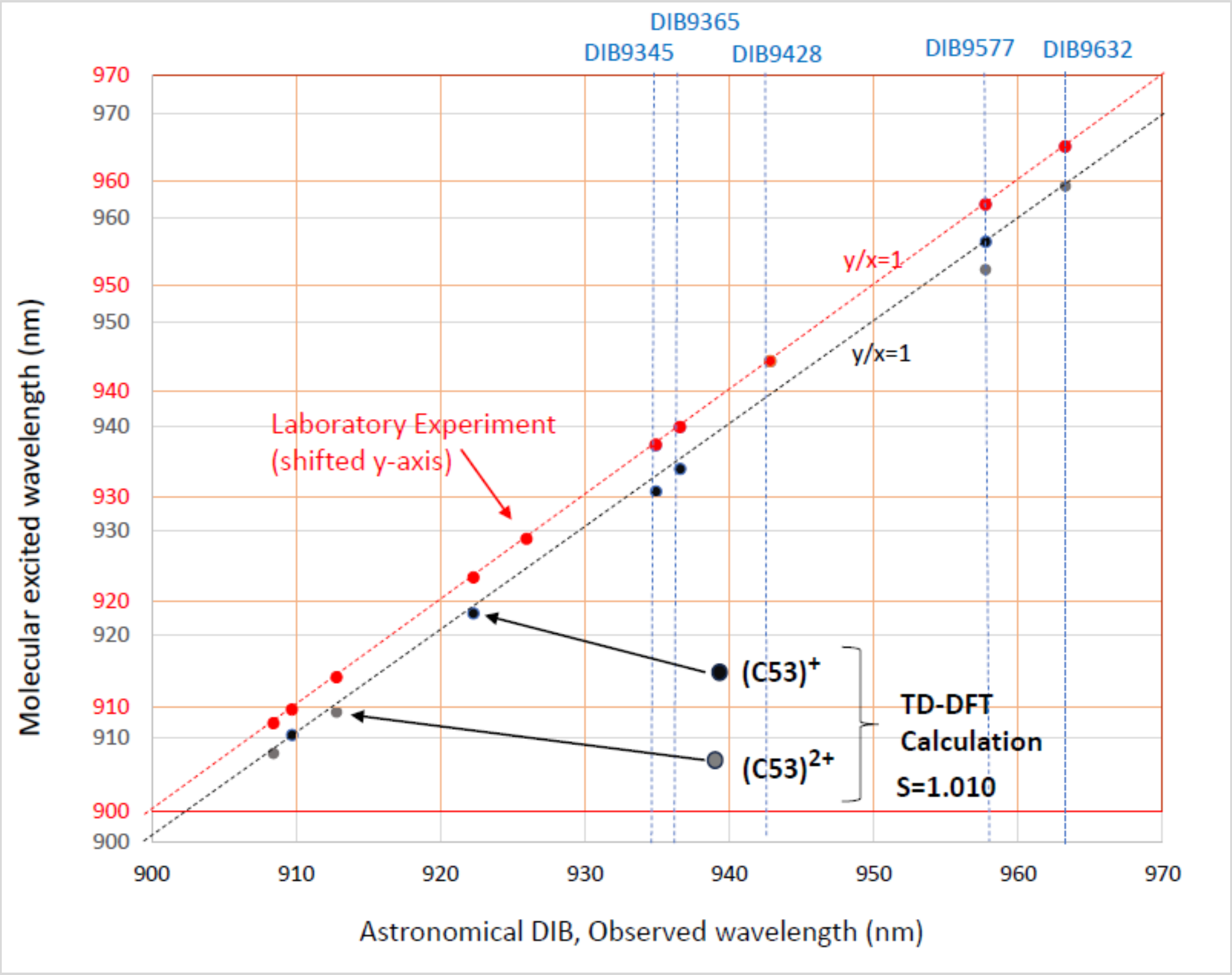

**Fig. 8** Graphene-($C_{53}$) family (s=1.010) shows good agreement both with astronomical DIBs (blue dotted vertical lines) and with laboratory experimental ones (red circles by shifted y-axis).

## 8. Conclusion

This study theoretically predicts specific pure carbon molecules to reproduce astronomical carbon related Diffuse Interstellar Bands (DIB).

(1) In 2015 and 2016, E. Cambell et al. opened laboratory experiments coincided with observed five DIB bands, which were explained to come from fullerene mono cation $(C_{60})^{1+}$.

(2) By Time-dependent DFT, molecular orbital excitation wavelength of $(C_{60})^{1+}$ was calculated. Calculated two bands were close to observed DIBs. However, there were serious theoretical problems that oscillator strength was zero and that other three DIBs could not be reproduced.

(3) Laboratory experiments was the mass spectroscopic one filtering m/e=724 molecule, which simply expect fullerene-$(C_{60})^{1+}$ combined with He. Other model molecules would be like He-atom intercalated graphite, as [graphene$(C_{53})^{1+}$--He--$(C_7)$], [graphene$(C_{51})^{1+}$ --He--$(C_9)$] and so on.

(4) A family of graphene-$(C_{53})$, $(C_{52})$, $(C_{51})$ was calculated. Results show that an astronomical 957.74nm band was reproduced well by calculated 957.74nm of $(C_{53})^{1+}$, also confirmed by laboratory experiment of 957.75nm. Other observed 963.26, 936.57 and 934.85nm bands were reproduced by calculated 963.08, 935.89 and 933.72nm.

(5) Experimentally measured 922.27nm band was calculated to be 922.02nm. Similarly, experimental 925.96, 912.80, 909.71 and 908.40nm bands were all reproduced by calculated 926.01, 912.52, 910.32 and 908.55nm.

It should be emphasized that graphene molecules may be ubiquitously floating in interstellar space.

## Acknowledgement

We would like to say great thanks to Prof. Aigen Li of Univ. Missouri for the important suggestions. We like to say special thanks to the excellent review paper by Dr. T.R. Geball.

---

Authors Profile : Norio Ota, PhD. (太田憲雄)
Magnetic materials and physics, Astrochemistry,
Honorary member of the Magnetics Society of Japan,
2010-2021: Senior Professor, Univ. of Tsukuba, Japan.
2003-2011: Exec. Chief Engineer, Hitachi Maxell Ltd.
https://www.researchgate.net/profile/Norio-Ota-Or-Ohta

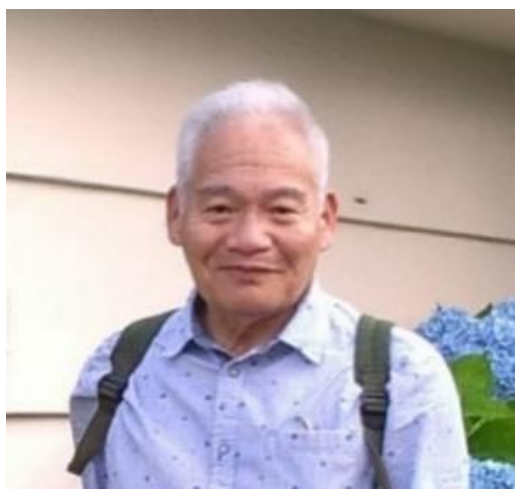

---